\documentclass[preprint]{revtex4}
\usepackage{color}
\usepackage{graphicx}
\usepackage{epsfig,amsmath,amssymb}
\newcommand{\be}{\begin{equation}}
\newcommand{\ee}{\end{equation}}
\newcommand{\ra}{\rangle}
\newcommand{\la}{\langle}
\newcommand{\bit}{\begin{itemize}}
\newcommand{\eit}{\end{itemize}}
\newcommand{\bea}{\begin{eqnarray}}
\newcommand{\eea}{\end{eqnarray}}
\newcommand{\Neel}{N\'{e}el}

\begin{document}
\title
{Ground-state properties of the triangular-lattice Heisenberg
antiferromagnet with arbitrary spin quantum number $s$}

\author
{O. G\"otze$^1$, J. Richter$^1$, R. Zinke$^2$ and D. J. J. Farnell$^3$\\
\small{$^1$ Institut f\"ur Theoretische Physik, Otto-von-Guericke-Universit\"at
Magdeburg,}\\ 
\small{P.O.B. 4120, 39016 Magdeburg, Germany}\\ 
\small{$^2$ Institut f\"ur Apparate- und Umwelttechnik,
Otto-von-Guericke-Universit\"at Magdeburg,}\\
\small{ P.O.B. 4120, 39016 Magdeburg, Germany}\\
\small{$^{3}$ 
   School of Dentistry, Cardiff University
   Cardiff CF14 4XY, Wales UK}
}

\begin{abstract}
We apply the coupled cluster method to high orders of approximation and
exact diagonalizations to study the 
ground-state properties of  
the triangular-lattice spin-$s$ Heisenberg antiferromagnet.
We calculate the fundamental ground-state quantities, namely, the 
energy $e_0$, the sublattice magnetization  $M_{\rm sub}$, the in-plane  spin
stiffness $\rho_s$ and 
the in-plane magnetic susceptibility $\chi$ for spin quantum numbers $s=1/2, 1,
\ldots,  s_{\rm max}$, where  $s_{\rm max}=9/2$ for  $e_0$ and $M_{\rm sub}$,
$s_{\rm max}=4$ for  $\rho_s$  and $s_{\rm max}=3$ for  $\chi$. 
We use the data  for $s \ge 3/2$ to
estimate the leading quantum corrections  to the classical values of $e_0$,
 $M_{\rm sub}$, $\rho_s$, and
$\chi$.
In addition, we study the magnetization process, the width of the 1/3
plateau  as well as the
sublattice magnetizations in the plateau state as a function of the spin quantum number
$s$.  
\end{abstract}

\date{\today}
\maketitle

\section{Introduction}

In the 1970s Anderson and Fazekas\cite{And,Faz}
first considered the quantum spin-$1/2$ Heisenberg antiferromagnet (HAFM)
for the geometrically
frustrated triangular
lattice and they proposed a liquid-like ground state (GS) without magnetic
long-range order (LRO).
Later on it was found
that the spin-$1/2$ HAFM on the triangular
lattice possesses semi-classical three-sublattice N\'{e}el order, see,
e.g.,
Refs.~\cite{jolicoeur1989,bernu1992,miyaka1992,chub94,bernu1994,bernu1995,manuel1998,capriotti1999,trumper00,ccm_previous,farnell2001,krueger2006,weihong2006,white2007,zhito2009,bishop2009,zhito2012}.
However, the sublattice magnetization $M_{\rm sub}$ is
drastically diminished in the $s=1/2$
model
\cite{capriotti1999,krueger2006,weihong2006,white2007,zhito2009,bishop2009}
because of the interplay between quantum fluctuations and strong frustration.
The small magnetic order parameter indicates that the semi-classical
magnetic LRO is fragile and that small additional terms in the
Hamiltonian may destroy the magnetic LRO, see, e.g.,
Refs.~\cite{cheng2011,serbyn2011,xu2012,bieri2012,extended1,extended2,extended3,extended4,extended5}.   

Although very precise data
for the relevant GS quantities are available for unfrustrated HAFM's on bipartite
two-dimensional lattices, see, e.g., 
Refs.~\cite{qmc1991,swt3rd,lin2001,ccm_square2007} related to  the square lattice, 
the corresponding data for the triangular
lattice are less precise. This lack of precision is related to the strong frustration
in the system  that, e.g., does not allow one to apply the quantum Monte Carlo
method.
Moreover, the spin-wave approach is less
efficient for frustrated lattices than it is for non-frustrated lattices.
Nevertheless, spin-wave theories are considered as appropriate, in
particular, if the spin  quantum number $s$ is not $s=1/2$ or $s=1$.
Perhaps the most accurate result for the GS order parameter (i.e., the
sublattice magnetization $M_{\rm sub}$) for $s=1/2$ has been obtained by
a recent density matrix renormalization group study \cite{white2007}, where
a result of $M_{\rm sub}= 0.205$ has been found.

The 
continuous interest in the triangular-lattice HAFM
is (last but not least) 
also related to a fluctuation-induced magnetization plateau at 1/3 of the
saturation
magnetization
\cite{kawa,nishi,chub_gol,Hon1999,ono2003,HSR04,farnell2009,alicea,tay2010,zhito2011,zhito2011a,shirata2011,shirata2012,wir_und_tanaka2013,satoshi2013,balents2013,tanaka2013a,chub2014,zhito2014,danshita2014,batista2015}.
In particular, two model compounds, namely Ba$_3$CoSb$_2$O$_9$
with $s=1/2$ and  
Ba$_3$NiSb$_2$O$_9$ with $s=1$, have been shown very recently to demonstrate
an excellent agreement
between the
experimentally measured magnetization curves and those curves from theoretical
predictions, see Refs.~\cite{farnell2009,shirata2012,tanaka2013a} for $s=1/2$ and
Refs.~\cite{shirata2011,wir_und_tanaka2013} for $s=1$.

In the present paper  we consider
the Hamiltonian
\begin{eqnarray}
\label{ham}
H=\sum_{\langle ij \rangle}{\bf s}_{i}{\bf s}_{j} - h \sum_i s_i^z ,
\end{eqnarray}
where the sum runs over nearest-neighbor bonds
${\langle ij \rangle}$ on the triangular lattice, $({\bf s}_{i})^2=s(s+1)$,
and $h$ is an external magnetic field.
We consider arbitrary spin quantum number $s$.
We use the coupled cluster method (CCM) to high orders of approximation
 to determine  the GS properties in zero magnetic field, i.e., the GS
energy per spin $e_0$, the sublattice magnetization  $M_{\rm sub}$  (order
parameter), the spin stiffness $\rho_s$, and the uniform susceptibility
$\chi$.
These quantities constitute 
the fundamental parameters determining  the low-energy physics of
the triangular Heisenberg antiferromagnet. Moreover,  the stiffness
and the susceptibility are used as input parameters in scaling functions for
various observables \cite{chub94_scal}.   
 
In addition to the zero-field quantities 
we also consider the magnetization process $M(h)$ and 
determine the $1/3$ plateau in the $M(h)$-curve.
We complement the CCM calculations by carrying out 
Lanczos exact diagonalization of finite lattices.

\begin{figure}[ht]%[ht]  h!
\begin{center}
\epsfig{file=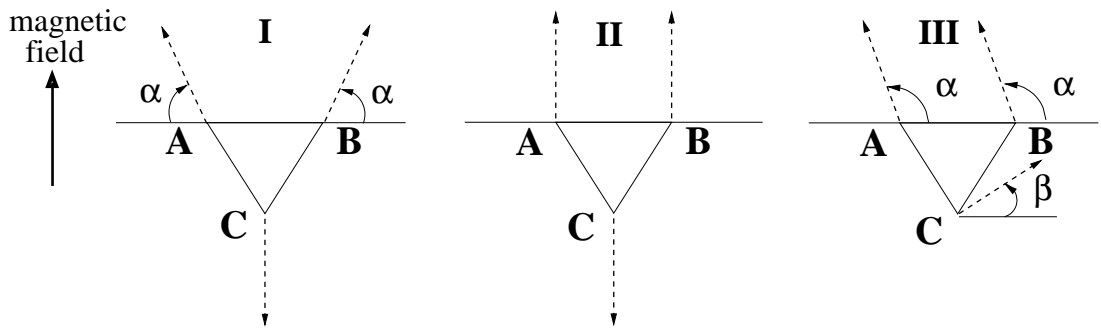 ,scale=0.55,angle=0.0}
\end{center}
\caption{Reference states used for the CCM calculations.
}
\label{fig1}
\end{figure}

\section{Methods}
\subsection{Lanczos exact diagonalization}
The Lanzcos exact diagonaliazion (ED) is one of the most useful methods that
can be used to
investigate frustrated quantum spin systems, see, e.g.,
Refs.~\cite{bccj1j2,star,ED40,Lauchli,lauchli2006,nakano2013,Sakai,capponi2013}.
Although lattices of size $N=36$ are common for ED calculations  for spin
$s=1/2$, 
the system size  $N$ accessible for  ED shrinks significantly, see, e.g.,
Refs.~\cite{hida2002,lauchli2006,nakano2013,Lauchli_s1,wir_und_tanaka2013,Sakai2015}. Hence, we use the ED here in order
to complement the results of the CCM (that yields results in the limit
$N\to\infty$).
We use  J. Schulenburg's {\it spinpack} code\cite{spinpack} to calculate the
magnetization curves for  $s=1/2,1,\ldots,5/2$.
The maximum lattice size for $s=2$ and $5/2$ is $N=12$, whereas for $s=3/2$ we
have results for $N=12,18,21$.  For $s=1$ the largest lattice we can consider
is $N=27$.
We use  these data to analyze the $s$-dependence of the $1/3$ plateau.

\subsection{Coupled cluster method} \label{ccm}
The coupled cluster method (CCM) is a universal many-body method widely used
in various fields of quantum many-body physics, see, e.g.
Refs.~\cite{bartlett97,bishop98}. 
Meanwhile, the CCM has been established as an effective tool in
the theory of frustrated quantum spin systems, see, e.g., 
the recent papers \cite{ccm_ferri,krueger2006,Schm:2006,rachid08,bishop08,ccm_odd_even,farnell2009,richter2010,farnell11,ccm_kago,wir_und_tanaka2013,trian_s1_ccm,archimedean2014,extended4,jiang2015,kago_xxz}. 
Here we illustrate only some features of
the CCM relevant for the present paper. 
For more general information on the methodology of the CCM, see, e.g.,
Refs.~\cite{ccm_roger1990,zeng98,bishop98,bishop00,farnell02,bishop04}.

The CCM
calculation starts with  the choice of a normalized reference  state
$|\Phi\rangle$. We choose the classical GS of the model as  reference  state,
which is well known   
for the triangular HAFM for arbitrary fields, see, e.g.,
Refs.~\cite{chub_gol,alicea,farnell2009} and Fig.~\ref{fig1}.
For zero field it is  three-sublattice {\Neel} state, i.e., state I with
$\alpha=60^o$ in Fig.~\ref{fig1}. For
finite magnetic fields non-collinear planar states with field dependent pitch angles
$\alpha$ and $\beta$
are classical GS's, see Fig.~\ref{fig1}. The
reference state is a collinear state (so-called up-up-down state, see state II in
Fig.~\ref{fig1})  only at the 1/3 plateau. 
With respect to the corresponding reference state, we then define a set of 
mutually commuting multispin
creation operators $C_I^+$, which are 
themselves defined over a complete set of
many-body configurations $I$.  
We perform a rotation of the local axis of
the spins such that all spins in the reference state align along the
negative $z$ axis.  
The specific form of the spin-operator transformation
 depends on the pitch angles of the
reference state.
In this new set of local spin coordinates 
the reference state and the corresponding multispin
creation operators $C_I^+$ are  given by
\begin{equation}
\label{set1} |{\hat \Phi}\ra = |\downarrow\downarrow\downarrow\cdots\rangle ; \mbox{ }
C_I^+ 
= {\hat s}_{n}^+ \, , \, {\hat s}_{n}^+{\hat s}_{m}^+ \, , \, {\hat s}_{n}^+{\hat s}_{m}^+{\hat
s}_{k}^+ \, , \, \ldots \; ,
\end{equation}
where the indices $n,m,k,\ldots$ denote arbitrary lattice sites.
In the rotated coordinate frame the Hamiltonian
becomes dependent on the pitch angles.
With the set $\{|\Phi\rangle, C_I^+\}$ the CCM parametrization of 
the exact ket and bra
GS eigenvectors 
$|\Psi\ra$ and $\la \tilde{ \Psi}|$ 
of the many-body system 
is given  by
\begin{eqnarray}
\label{eq5} 
|\Psi\ra=e^S|\Phi\ra \; , \mbox{ } S=\sum_{I\neq 0}a_IC_I^+ \; \\
\label{eq5b}
\la \tilde{ \Psi}|=\la \Phi |\tilde{S}e^{-S} \; , \mbox{ } \tilde{S}=1+
\sum_{I\neq 0}\tilde{a}_IC_I^{-} \; ,
\end{eqnarray}
where $C_I^-=\left (C_I^+ \right )^{\dagger}$.
The CCM correlation operators, $S$ and $\tilde{S}$, contain the correlation
coefficients,
$a_I$ and $\tilde{a}_I$, 
which can be determined by the CCM ket-state
and bra-state
equations
\begin{eqnarray}
\label{eq6}
\langle\Phi|C_I^-e^{-S}He^S|\Phi\rangle = 0 \;\; ; \; \forall I\neq 0  \\
\label{eq6a}\langle\Phi|{\tilde S}e^{-S}[H, C_I^+]e^S|\Phi\rangle = 0 \; \;
; \; \forall
I\neq 0 .
\end{eqnarray}
Note that each ket-state 
equation belongs to a specific creation operator\\
$C_I^+=s_n^+,\,\,s_n^+s_{m}^+,\,\, s_n^+s_{m}^+s_{k}^+,\cdots$,
i.e., it corresponds to a specific  set (configuration) of lattice sites
$n,m,k,\dots\;.$ 
By using the Schr\"odinger equation, $H|\Psi\ra=E|\Psi\ra$,
we can write the GS energy as $E=\la\Phi|e^{-S}He^S|\Phi\ra $. 
The sublattice magnetization is given by 
 $M_{\rm sub}  = -(1/N) \sum_i^N \la\tilde\Psi|s_i^z|\Psi\ra$, where $s_i^z$ is
expressed in the transformed coordinate system.
The total magnetization $M$ 
aligned in
the direction of the applied magnetic field
$h$   
in terms of the global axes prior to rotation of the local spin axes
is given by  $M=(M_A+M_B+M_C)/3$, where $M_A$, $M_B$, and $M_C$ are
the magnetizations of the three individual sublattices, cf. Fig.~\ref{fig1}, given by 
\begin{equation}
M_{A,B,C} =
   \frac 1{N_{A,B,C}}   \sum_{i_{A,B,C}}
      \langle \tilde \Psi | s_{i_{A,B,C}}^z | \Psi \rangle ,
 \label{rotMabc}
\end{equation}
where the index $i_A$ runs over all $N_A$ sites on sublattice $A$,
the index $i_B$ runs over all $N_B$ sites on sublattice $B$, and the
index $i_C$ runs over all $N_C$ sites on sublattice $C$, and
$N=N_A+N_B+N_C$.
The CCM results for the ground state energy and the total
magnetization as a function of the magnetic field can be used 
to 
calculate the uniform magnetic susceptibility, given by
\begin{equation} \label{susc}
\chi \equiv \frac{dM}{d h} = 
- \frac{1}{N}\frac{d^2 E}{d h^2} \; .
\end{equation}
Note that we consider here $\chi$ as susceptibility per site
\cite{defin_chi}.

The GS energy depends
(in a certain CCM approximation, see below)
on the pitch angles.
In the quantum model the pitch angles may be different to the corresponding classical
values. 
Therefore, we do not choose the classical result for the pitch angles in the quantum
model.
Indeed, we consider them as a free
parameter in the CCM calculation, which has to be determined by minimization of the
CCM GS energy with respect to the pitch angles.
An exception is the zero-field case, where  
the pitch angle 
is fixed to $\alpha=60^o$ (the three-sublattice N\'{e}el state).

The spin stiffness $\rho_s$ 
measures the increase of energy rotating the  
order parameter of a magnetically long-range ordered system along a given
direction by a small twist (pitch) angle
$\theta$ per unit length, i.e., 
\be\label{stiffn}
\frac{E(\theta)}{N} = \frac{E(\theta=0)}{N} + \frac{1}{2}\rho_s \theta^2 +
{\cal O}(\theta^4) ,
\ee
where $E(\theta)$ is the ground-state energy as a function of the twist
angle. For the triangular lattice the twist is imposed along a lattice basis
vector and it is within the plane defined by the order
parameter, see Refs.~\cite{bernu1995,manuel1998}, where
the twist along  both directions leads to identical
results \cite{comment_twist}.

For the many-body quantum system under consideration
it is necessary to use approximation 
schemes in order to truncate the expansions of $S$ and $\tilde S$
in  Eqs.~(\ref{eq5}) and (\ref{eq5b}) in a practical calculation.
We use the well established SUB$n$-$n$ approximation scheme, cf., e.g.,
Refs.~\cite{farnell02,bishop04,Schm:2006,krueger2006,rachid08,bishop08,ccm_odd_even,farnell2009,richter2010,farnell11,ccm_kago,wir_und_tanaka2013,trian_s1_ccm,archimedean2014,extended4,jiang2015,kago_xxz,zeng98,bishop00}, 
where the correlation operators   
contain no more than $n$ spin flips spanning a range of no more than
$n$ contiguous lattice sites \cite{SUBn-n}.

Using an efficient 
parallelized CCM code \cite{cccm} we are able to solve the CCM equations up
to SUB10-10 for $s=1/2$ (where, e.g., for the zero-field case
reference state a set of 1\;054\;841 coupled ket-state equations has to be
solved).
For $s>1/2$ the number of  CCM equations increases noticeably. Hence, the highest order of approximation is
then SUB8-8 (where, e.g., for the susceptibility for $s=3$ a set of 2\;179\;007 equations has to be
solved).

The SUB$n$-$n$ approximation becomes exact only for $n \to \infty$.
We extrapolate the `raw' SUB$n$-$n$
data to $n \to \infty$. 
Much experience exists relating to the 
extrapolation of the GS energy per site $e_0(n) \equiv E(n)/N$, the magnetic order parameter
 $M_{\rm sub}$, the spin stiffness $\rho_s$, and the susceptibility $\chi$.
Thus  $e_0(n) = a_0 + a_1(1/n)^2 + a_2(1/n)^4$ is a very
well-tested extrapolation
ansatz for the GS energy per spin
\cite{bishop00,bishop04,Schm:2006,bishop08,ccm_odd_even,rachid08,archimedean2014}.  An appropriate extrapolation rule
for the magnetic order parameter of antiferomagnets with GS LRO
is 
$M_{\rm sub}=b_0+b_1(1/n)+b_2(1/n)^{2}$ \cite{bishop00,bishop04,farnell2009,trian_s1_ccm,archimedean2014}.
For the stiffness $\rho_s$ as well as for the susceptibility $\chi$ 
we use the same rule as for $M_{\rm sub}$, i.e., $X(n)=c_0+c_1(1/n)+c_2(1/n)^{2}$,
$X=\rho_s, \chi$, which is able to describe the asymptotic behavior of the 
CCM-SUB$n$-$n$ data for $\rho_s$ well,  see  Refs.~\cite{krueger2006,rachid08},
and $\chi$, see Ref.~\cite{farnell2009}.

The selection of the SUB$n$-$n$ data included in the extrapolation is a
subtle issue. Often it is argued that the lowest-order data (i.e.,  SUB2-2 and SUB3-3)
ought to be excluded from the extrapolation because  
these points are rather far from the asymptotic
regime \cite{ccm_kago,archimedean2014,extended4}.
This argument is particularly valid for  models which include
larger-distance  exchange
bonds (e.g., so-called $J_1$-$J_2$ models)
\cite{rachid08,bishop08,richter2010,farnell11,extended4}.
However, for the triangular Heisenberg antiferromagnet with only
nearest-neighbor bonds even the lowest approximation orders fit well to the
extrapolation \cite{krueger2006,farnell2009,wir_und_tanaka2013}.   
Another point is the odd-even problem, i.e., for odd
and even numbers $n$ of the SUB$n$-$n$ approximation the extrapolation may
have different fit parameters \cite{ccm_odd_even,archimedean2014}.
However, this problem occurs primarily for bipartite systems (with collinear
reference states, where no odd-numbered spin flips enter the 
correlation operators $S$ and $\tilde{S}$), whereas for  noncollinear
reference states (where odd-numbered spin flips are present in  $S$ and
$\tilde{S}$) relevant for 
many frustrated systems both, odd and
even  SUB$n$-$n$, might be combined in one and the same extrapolation
formula \cite{krueger2006,ccm_kago,archimedean2014}.

\begin{figure}[ht]%[ht]  h!
\begin{center}
\epsfig{file=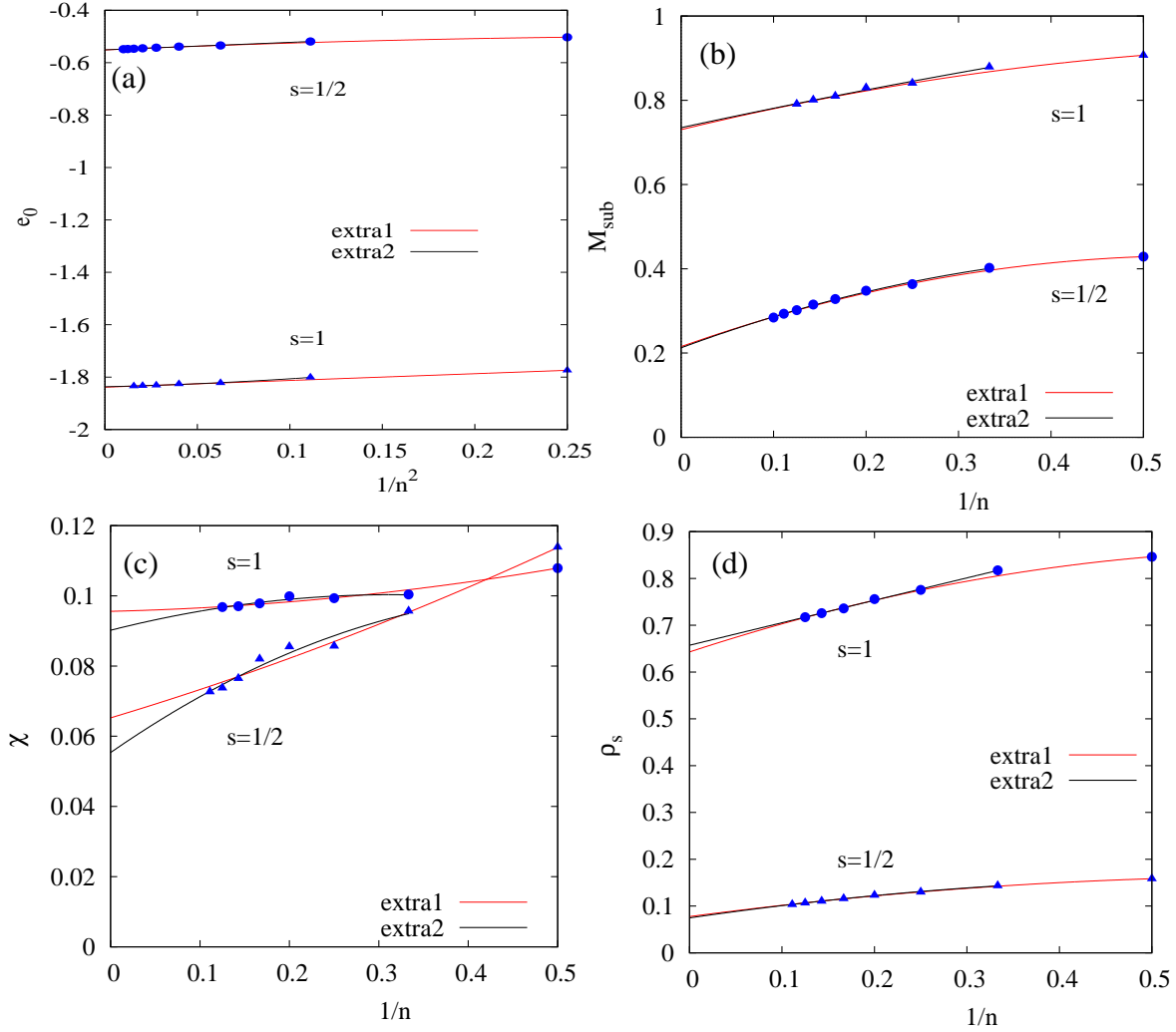,scale=0.8,angle=0.0}
\end{center}
\caption{Extrapolation of the CCM-SUB$n$-$n$ for the 
GS energy $e_0$ (a), the sublattice magnetization  $M_{\rm sub}$ (b), the
susceptibility $\chi$ (c), and the spin
stiffness $\rho_s$ (d) for spin quantum numbers $s=1/2$ 
and $1$ using two different extrapolation schemes (labeled by 'extra1' and
'extra2'),
cf. main text.
}
\label{extra_s0.5_s1}
\end{figure}

In order to fit the data to
the extrapolation formulas given above (which contain three unknown
parameters), it is desirable (as a rule) 
to have at least four data points to obtain a robust and stable fit.
To obey this rule we apply here the above given extrapolation formulas using 
(i) even $n=2,4,6,\ldots$  and
(ii)  odd and even $n=3,4,5,6,7,\ldots$\cite{comp_archi}.
The maximum approximation level SUB$n_{\rm max}$-$n_{\rm max}$ for $s=1/2$ is $n_{\rm max}=10$ for
$e_0$ and $M_{\rm sub}$ and $n_{\rm max}=9$ for    
$\chi$ and $\rho_s$, whereas for $s>1/2$ we have  $n_{\rm max}=8$.
In the following  we call case (i) 'extra1' and  case (ii) 'extra2'.
The difference between both cases can be considered as a measure of accuracy
of our CCM results.
To illustrate our extrapolation procedure, we present in
Fig.~\ref{extra_s0.5_s1} the extrapolations of 
$e_0$, $M_{\rm sub}$, $\chi$, and $\rho_s$ for $s=1/2$ and $s=1$.
Obviously,
the extrapolations work very well for $e_0$, $M_{\rm sub}$, and
$\rho_s$, and, both schemes, 'extra1' and 'extra2', lead to very
similar results.
 There is some scattering of the  SUB$n$-$n$ data only for $\chi$, and, as a consequence,
there is a visible difference between the extrapolations 'extra1' and
'extra2'.

\begin{table}[htb]
     \centering
\caption{Extrapolated CCM results for the GS energy per spin, $e_0 \vert_{n 
\to \infty}$,
the GS sublattice magnetization, $M_{\rm sub}\vert_{n \to \infty}$, the spin
stiffness, $\rho_s\vert_{n \to
 \infty}$, and the susceptibility, $\chi\vert_{n \to \infty}$.
We mention that the spin-wave velocity $c_{\rm swt}$ can be calculated from
 $\rho_s$ and $\chi$ by using the hydrodynamic relation $c^2_{\rm  swt}=\rho_s/\chi$.   
}
$ $\\
     \begin{tabular}{lrrrrrrrrr}\hline\hline
     \parbox[0pt][1.5em][c]{0cm}{}        & \multicolumn{4}{c} {extra1} &
  \multicolumn{4}{c}{extra2} \\\hline
       &\; $e_0/s^2$ &\; $M_{\rm sub}/s$ &\; $\rho_s/s^2$ &\; $\chi$ \; &\; 
$e_0/s^2$ &\; $M_{\rm sub}/s$&\; $\rho_s/s^2$ &\; $\chi$ \;\\ \hline
     $s=1/2$           &\; -2.2056   &\;  0.4307 &\; 0.3103 &\;  0.0652 
&\; -2.2045  &\;  0.4248 &\; 0.2990 &\;  0.0553 \\
     $s=1$             &\; -1.8384   &\;  0.7303 &\; 0.6429 &\;  0.0956 
&\; -1.8367  &\;  0.7350 &\; 0.6572 &\;  0.0902 \\
     $s=3/2$           &\; -1.7234   &\;  0.8169 &\; 0.7636 &\;  0.0996 
&\; -1.7223  &\;  0.8232 &\; 0.7757 &\;  0.0972 \\
     $s=2$             &\; -1.6667   &\;  0.8628 &\; 0.8246 &\;  0.1023 
&\; -1.6659  &\;  0.8695 &\; 0.8342 &\;  0.1012 \\
     $s=5/2$           &\; -1.6329   &\;  0.8909 &\; 0.8610 &\;  0.1041 
&\; -1.6323  &\;  0.8973 &\; 0.8687 &\;  0.1035 \\
     $s=3$             &\; -1.6105   &\;  0.9096 &\; 0.8850 &\;  0.1054 
&\; -1.6000  &\;  0.9155 &\; 0.8913 &\;  0.1050 \\
     $s=7/2$           &\; -1.5946   &\;  0.9229 &\; 0.9019 &\;- \;\;\  
&\; -1.5941  &\;  0.9282 &\; 0.9071 &\;- \;\;\, \\
     $s=4$             &\; -1.5826   &\;  0.9328 &\; 0.9145 &\;- \;\;\  
&\; -1.5823  &\;  0.9376 &\; 0.9188 &\;- \;\;\, \\
     $s=9/2$           &\; -1.5734   &\;  0.9404 &\;- \;\;\, &\;- \;\;\  
&\; -1.5731  &\;  0.9448 &\;- \;\;\, &\;- \;\;\,
     \\\hline \hline
  \end{tabular}
\label{table1}
\end{table}

\begin{figure}[ht!]%[ht]  h!
\begin{center}
\epsfig{file=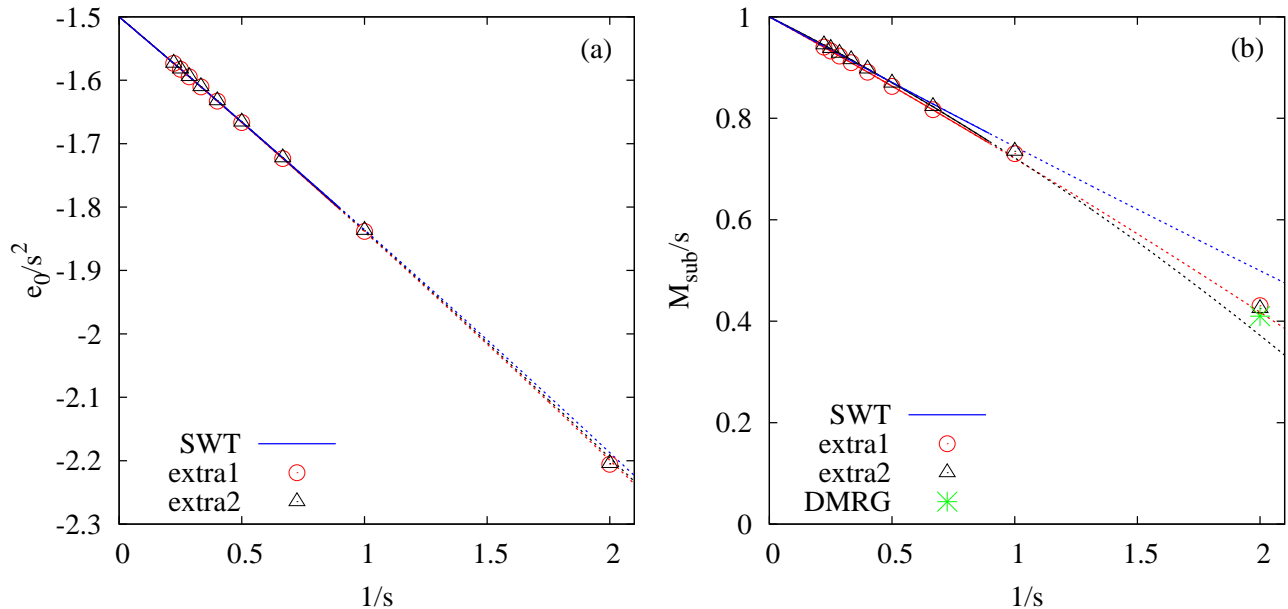,scale=0.95,angle=0.0}
\end{center}
\caption{
Extrapolated CCM data (schemes 'extra1' and 'extra2',
see main text) as a function of  $1/s$ compared with higher-order spin-wave theory (SWT, blue line) taken from
Ref.\cite{zhito2009}. The symbols represent the CCM data points, the corresponding red
and black lines show the fit function
$X(s)=X|_{s\to\infty}\left(1-x_1/s-x_2/s^2\right)$, where  
the data for  $s \ge 3/2$ were used for the fit.     
(a): GS energy $e_0$.
(b) Sublattice magnetization $M_{\rm sub}$ (the green symbol
shows the density matrix renormalization group result (DMRG) for $s=1/2$
from Ref. \cite{white2007}).
}
\label{E_M_vs_s}
\end{figure}        

\begin{table}[ht!]
    \centering
\caption{Parameters $x_1$ and $x_2$ of the $1/s$ expansion
 $X(s)=X|_{s\to\infty}\left(1-x_1/s-x_2/s^2\right)$  obtained from the extrapolated CCM results 
for the GS energy $e_0$, the sublattice magnetization $M_{\rm sub}$, the
 spin stiffness $\rho_s$ and the susceptibility $\chi$.  
}
$ $\\
    \begin{tabular}{lrrrr}\hline\hline
    \parbox[0pt][1.5em][c]{0cm}{}        & \multicolumn{2}{c} {$e_0$} &  $M_{\rm sub}$ \\\hline
      & \; $x_1$ &\; $x_2$ &\; $x_1$&\; $x_2$\\ \hline
    extra1           &\; 0.2176     &\;  0.0071   &\; -0.2671   &\; -0.0120  \\
    extra2             &\; 0.2186   &\;  0.0073 &\; -0.2416   &\;    -0.0362     
    \\\hline \hline
    \parbox[0pt][1.5em][c]{0cm}{}        & \multicolumn{2}{c} {$\rho_s$} & $\chi$ \\\hline
      & \; $x_1$ &\; $x_2$ &\; $x_1$&\; $x_2$\\ \hline
    extra1           &\; -0.3355     &\;  -0.0292    &\; -0.1587   &\;  0.0045  \\
    extra2           &\; -0.3166     &\;  -0.0297    &\; -0.1440   &\;  -0.0662     
    \\\hline \hline
 \end{tabular}
\label{table2}
\end{table}
\begin{figure}[ht!]%[ht]  h!
\begin{center}
\epsfig{file=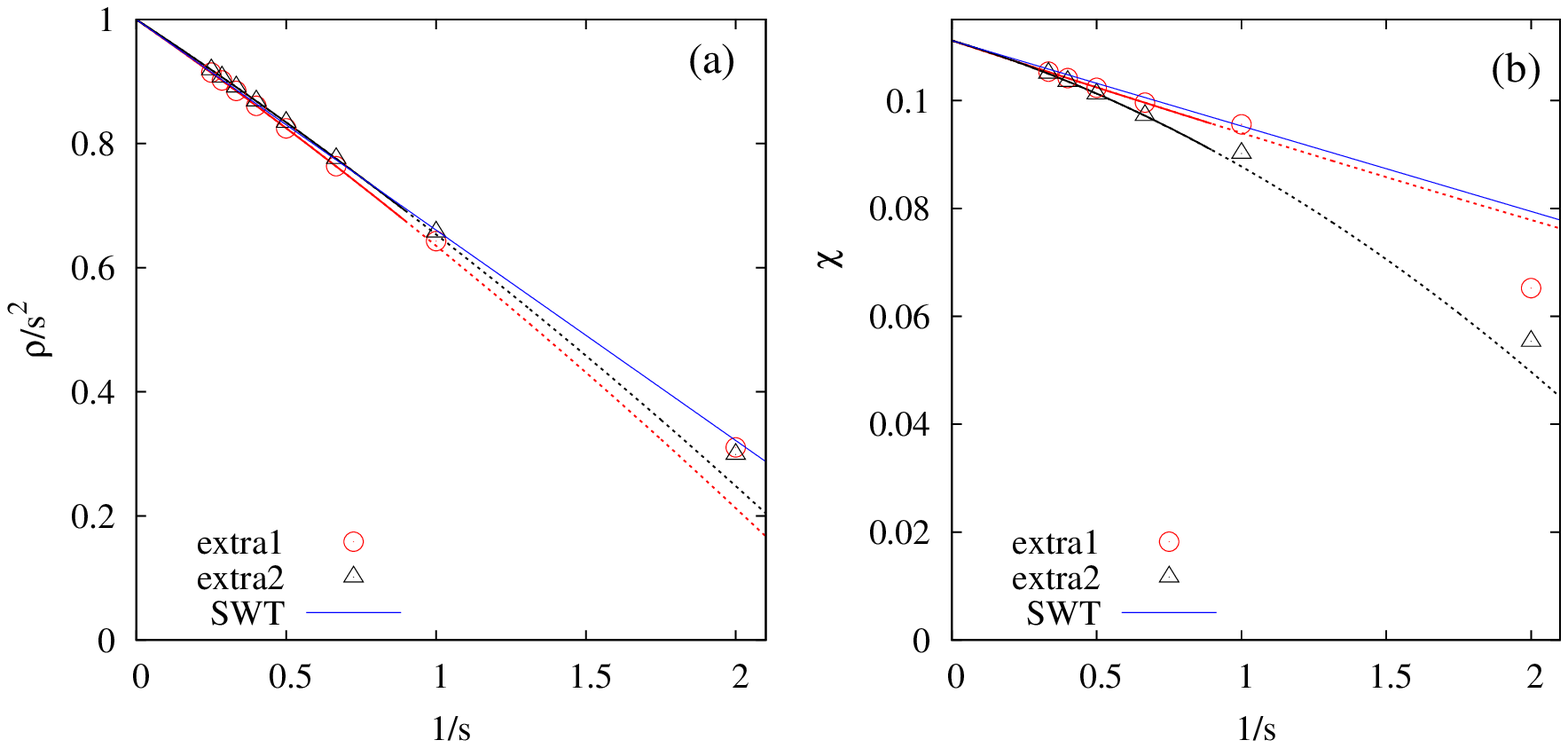,scale=0.95,angle=0.0}
\end{center}
\caption{
Extrapolated CCM data  (schemes 'extra1' and 'extra2',
see main text) for the spin stiffness $\rho_s$ (a)  and the susceptibility $\chi$
(b)
as a function of  $1/s$ compared with spin-wave theory (SWT, blue line) taken 
from
Ref.~\cite{bernu1995} (for  $\rho_s$) and Ref.~\cite{chub94} (for $\chi$). 
The symbols represent the CCM data points, the corresponding red
and black lines show the fit function
$X(s)=X|_{s\to\infty}\left(1-x_1/s-x_2/s^2\right)$, where for the fit 
the data for  $s \ge 3/2$  were used.     
}
\label{rho_chi_vs_s}
\end{figure}

\section{Results}
\subsection{The zero-field case}
\label{H_0}
In this section we present CCM results for the GS
energy per spin $e_0$, the sublattice magnetization $M_{\rm sub}$ (order
parameter), the spin stiffness $\rho_s$, and the uniform susceptibility
$\chi$ for spin quantum numbers $s=1/2, 1, \ldots, 9/2$ (for $e_0$ and $M_{\rm
sub}$), for $s=1/2, 1, \ldots, 4$ (for $\rho_s$ ), and for $s=1/2, 1,
\ldots, 3$ (for $\chi$).
Moreover, we use the data for
$s \ge 3/2$ to estimate the leading
quantum corrections to the classical values to compare with the $1/s$
spin-wave expansion \cite{bernu1995,miyaka1992,chub94,zhito2009}.

The data for the GS
energy and  the sublattice magnetization  are collected in
Table~\ref{table1}. The difference between both extrapolation schemes,
'extra1'
and 'extra2', is largest for lower spin quantum numbers $s$, although it is
still small for all values of $s$.
In the extreme quantum limit $s=1/2$ the density matrix
renormalization group result \cite{white2007}
$M_{\rm sub}/s= 0.410$ is slightly lower than our CCM result.

Let us now compare our data for  $e_0$ and $M_{\rm sub}$
with recent higher-order spin-wave
results by Chernyshev and Zhitomirsky \cite{zhito2009}.
 Chernyshev and Zhitomirsky found that 
$e_0(s)=-1.5s^2(1 + 0.218412/s + 0.0053525/s^2)$
and  $M_{\rm sub}(s) = s(1 - 0.261 303/s + 0.0055225/s^2)$.
We fit our extrapolated CCM data for $s=3/2,2,\ldots,9/2$
using  the ansatz\\
$X(s)=X|_{s\to\infty}\left(1-x_1/s-x_2/s^2\right)$, $X=e_0,M_{\rm sub}$. 
The classical values are $e_0|_{s\to\infty}=-3s^2/2$,
and $M_{\rm sub}|_{s\to\infty}=s$.

The values for the $1/s$ expansion parameters $x_1$ and $x_2$ are listed in
Table~\ref{table2} and   the corresponding results are depicted in Fig.~\ref{E_M_vs_s}. 
For the GS energy $x_1$ and $x_2$ are in very good agreement with the
spin-wave results  \cite{zhito2009}.  The leading
coefficient $x_1$ for the order parameter also fits well to the spin-wave
term. However,  we obtain (small) negative values instead of a
(small) positive one for the
next-order coefficient $x_2$.   
The good agreement  between the spin-wave and the CCM results for the GS energy is
also evident in Fig.~\ref{E_M_vs_s}(a).
Moreover, the $1/s$ expansion up to second order yields reasonable results for $e_0$
even for the extreme quantum case $s=1/2$.      
On the other hand, the deviation for the  sublattice magnetization becomes noticeable for
$s<3/2$, see  Fig.~\ref{E_M_vs_s}(b). Thus, by contrast to the GS energy, the $1/s$ expansion of $M_{\rm sub}/s$ up to order $s^{-2}$ 
leads to values for $s=1/2$ with limited accuracy.
We know that  the sublattice magnetization 
of the unfrustrated square-lattice $s=1/2$ Heisenberg
antiferromagnet obtained by higher-order spin-wave theory\cite{swt3rd} agrees
well with quantum Monte Carlo\cite{qmc1991} and CCM\cite{ccm_square2007}
results. Thus the deviation for the triangular lattice      
might be attributed to the enhanced quantum fluctuations caused by
frustration leading to a particularly small value of $M_{\rm sub}$ for $s=1/2$.

Next we discuss the in-plane spin
stiffness $\rho_s$ and
the in-plane magnetic susceptibility $\chi$. Results are given in Table~\ref{table1}. 
As already mentioned above, the
extrapolation of the CCM-SUB$n$-$n$ data  works well for $\rho_s$, although it is
less accurate for $\chi$, i.e.  for $\chi$ the deviation between the schemes 'extra1' and
'extra2' is noticeable, cf. Table~\ref{table1} and Fig.~\ref{extra_s0.5_s1}.
The spin-wave large-$s$ relations are  $\rho_s=s^2(1-0.3392/s)$
(Ref.~\cite{bernu1995}) and
$\chi=(1/9)(1-0.1425/s)$ (Ref.~\cite{chub94}).       
Fitting the data from Table~\ref{table1} for $s\ge 3/2$
using  the ansatz
$X(s)=X|_{s\to\infty}\left(1-x_1/s-x_2/s^2\right)$, $X=\rho_s, \chi$, 
$\rho_s|_{s\to\infty}=s^2$,
and $\chi|_{s\to\infty}=1/9$, gives values for the $1/s$ expansion
parameters $x_1$ and $x_2$ which are listed in Table~\ref{table2}.
The leading
coefficient $x_1$ fits reasonably well to the spin-wave term.    
The deviation between the schemes 'extra1' and
'extra2' for $\chi$ results in different signs of the second-order term
$x_2$.
We show the $1/s$ dependence of $\rho_s$ and $\chi$ in
Fig.~\ref{rho_chi_vs_s}. 
It is obvious that the large-$s$ approach works surprisingly well for  $\rho_s$ down to
$s=1/2$, whereas it seems to fail for $\chi$ for $s=1/2$.

\begin{figure}[ht!]%[ht]  h!
\begin{center}
\epsfig{file=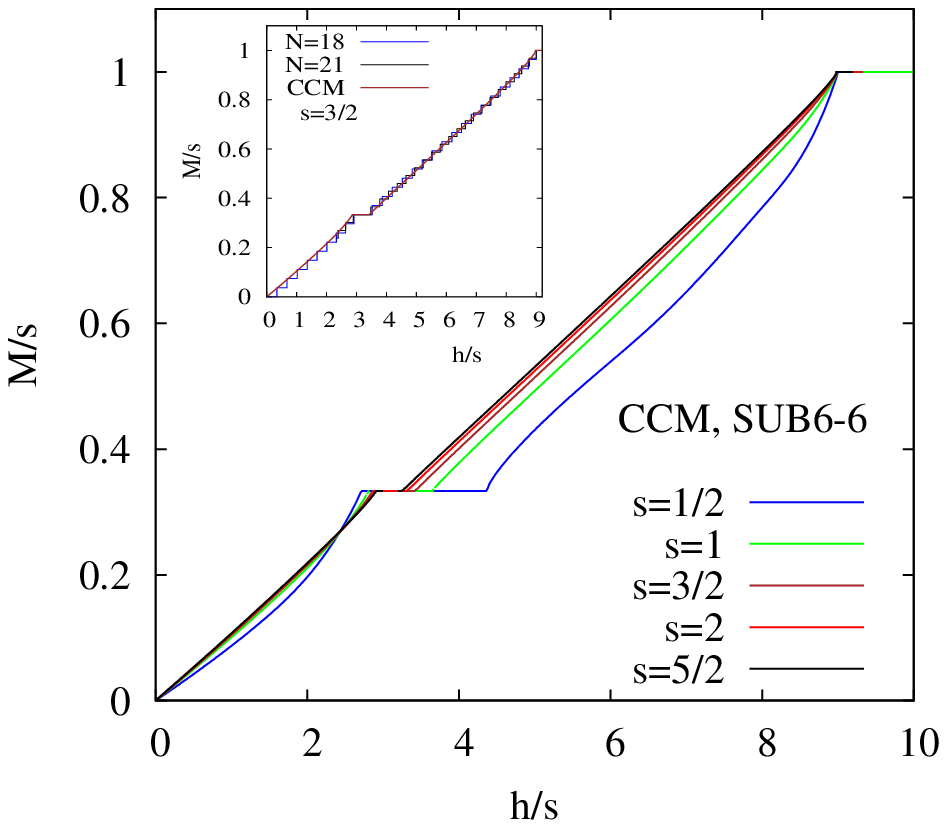,scale=1.1,angle=0.0}
\end{center}
\caption{Main: The magnetization curve calculated within CCM-SUB6-6 approximation
for $s=1/2,1,3/2,2$ and $5/2$. Inset:
The magnetization curve for $s=3/2$ calculated  with exact diagonalization
for finite lattices of $N=18$ and $21$ sites compared with the CCM-SUB6-6 data.
}
\label{m_h}
\end{figure}

\subsection{The magnetization process}

As already mentioned in the introduction, the magnetization process $M(h)$ for
$s=1/2$ was investigated previously in numerous
papers
\cite{nishi,chub_gol,Hon1999,HSR04,farnell2009,tay2010,zhito2011,shirata2012,satoshi2013,balents2013,tanaka2013a,zhito2014,danshita2014,batista2015}.
The $M(h)$ curve for the specific case $s=1$ was much less
studied \cite{nakano2013,wir_und_tanaka2013}.
Several
quasi-classical large-$s$ approaches\cite{chub_gol,alicea,zhito2011,zhito2014,chub2014}  
can be used to obtain an estimate for $M(h)$  and the $1/3$ plateau also for lower spin
quantum numbers. However, it is likely that these results have limited accuracy (see also the discussion
in Sec.~\ref{H_0}).
We mention again that very good agreement between
experimental and theoretical CCM data for
$s=1/2$ and $s=1$ has been reported \cite{shirata2012,wir_und_tanaka2013}
very recently. 

Let us mention that the CCM calculations of the $M(h)$ curves are extremely time
consuming, because  the field dependent
quantum pitch angles for each value of the magnetic field have to be
determined by minimization of the
CCM-SUB$n$-$n$  GS energy with respect to the pitch angles. 
Hence we consider only even SUB$n$-$n$ approximations until $n=6$.  
 We have data for SUB8-8 only for the critical fields, $h_{c1}$ and $h_{c2}$, which bound  the  $1/3$
plateau, and only for the most interesting extreme quantum cases of $s=1$ and $s=1/2$.
We show the CCM-SUB6-6 magnetization curves for $s=1/2, 1, 3/2, 2$, and $5/2$ in
the main panel of Fig.~\ref{m_h}. The width of the $1/3$ plateau shrinks with
increasing spin quantum number  $s$ which is clearly
seen in Fig.~\ref{m_h}. From the experimental point of view this shrinking of the plateau width 
is relevant. Thus for $s=2$ the plateau width $(h_{c2}-h_{c1})/s$ is only about   
$25\%$ of the width for $s=1/2$ which makes its detection for large $s$ by
measurements at
low (but finite) temperatures more challenging.
From  Fig.~\ref{m_h} it is obvious that 
all curves for $s>1/2$ are close to each other. Below and above the
plateau they show  almost the classical linear $h$ dependence of $M$.
The   $s=1/2$ curve is well 
separated and shows a pronounced deviation from linearity.   
The curves shown in the inset demonstrate that CCM and ED data agree well.

\begin{figure}[ht!]%[ht]  h!
\begin{center}
\epsfig{file=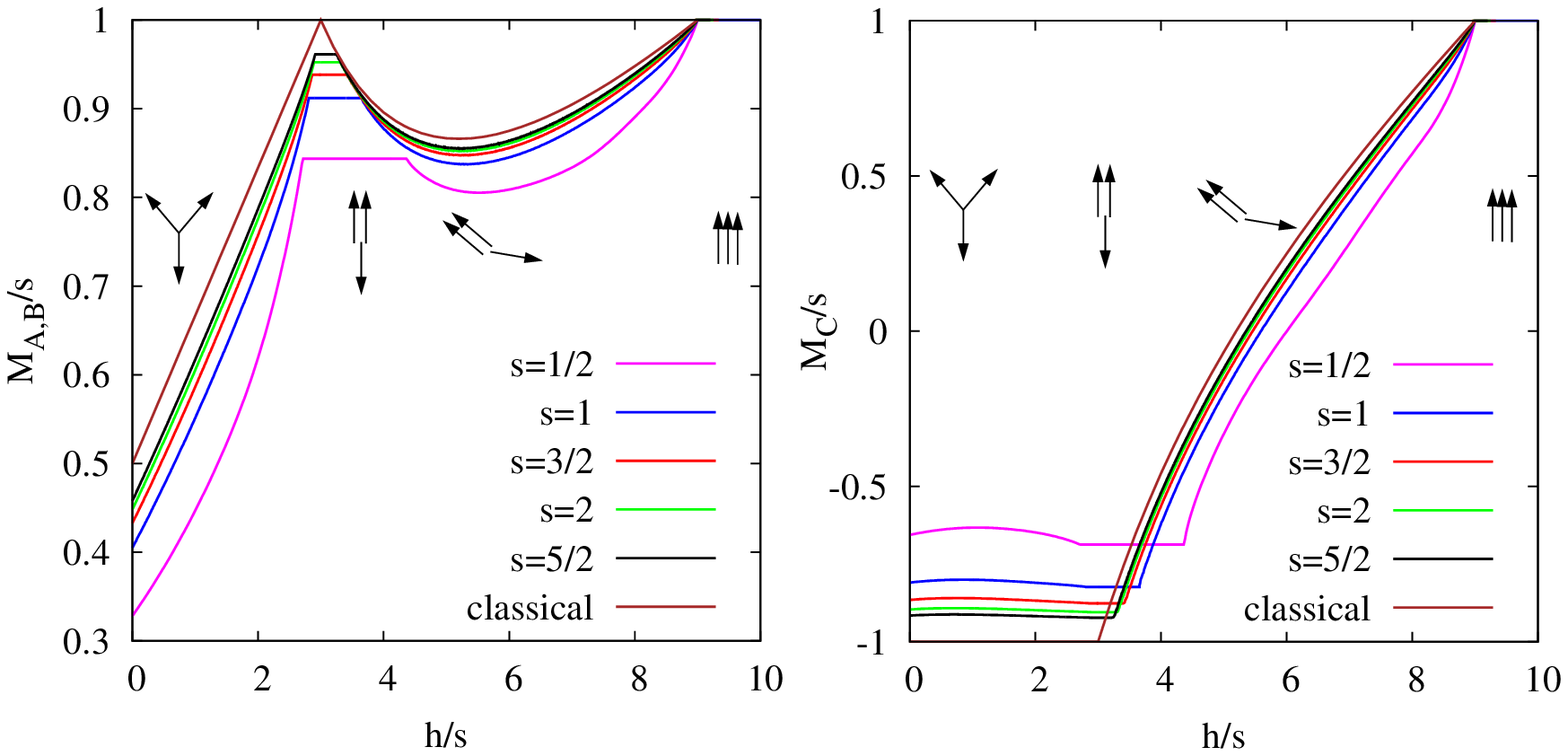,scale=0.8,angle=0.0}
\end{center}
\caption{Field dependence of the sublattice magnetizations $M_A=M_B$ (left)
and $M_C$ (right), cf. Fig.~\ref{fig1}, calculated within CCM-SUB6-6 approximation
for $s=1/2,1,3/2,2$ and $5/2$. 
}
\label{M_ABC}
\end{figure}  
\begin{figure}[ht!]%[ht]  h!
\begin{center}
\epsfig{file=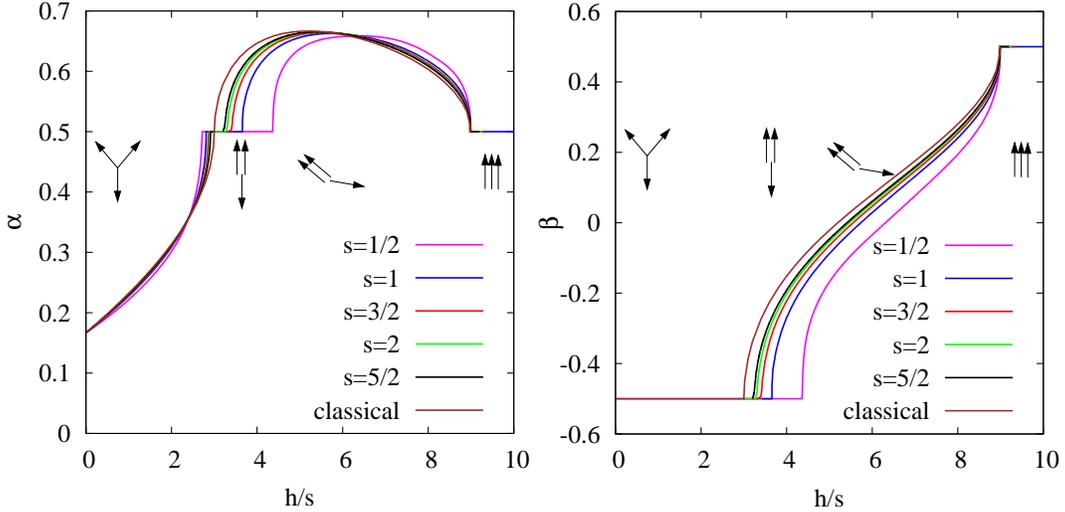,scale=0.8,angle=0.0}
\end{center}
\caption{Field dependence of the pitch angles $\alpha$ (left)
and $\beta$ (right), cf. Fig.~\ref{fig1}, calculated within CCM-SUB6-6
approximation
for $s=1/2,1,3/2,2$ and $5/2$.
}
\label{angles}
\end{figure}

The CCM approach allows to calculate the individual sublattice
magnetizations $M_A=M_B$ and $M_C$ as well as the quantum pitch angles
$\alpha$ and $\beta$, cf. Fig.~\ref{fig1}.
These quantities, are accessible, in principle, in neutron scattering
experiments. They provide a deeper insight in the details of the magnetization
process and the role of quantum fluctuations.   
We show  $M_A,M_B$ and $M_C$  in Fig.~\ref{M_ABC} and $\alpha$ and $\beta$
in Fig.~\ref{angles}. An interesting feature is the non-monotonic behavior of
$M_C$ for $h< h_{c1}$ (present only in the quantum model) and of 
$M_A,M_B$ and $\alpha$ above the plateau.  
There is a strong
increase in the slopes (except for $M_C$ and $\beta$ near $h_{c1}$) as one 
one approaches the plateau from below or above. 

For the collinear plateau state at one third of the saturation  
(the so-called 'up-up-down' state, see state II in
Fig.~\ref{fig1}) we have calculated SUB$n$-$n$ data of the sublattice
magnetizations $M_{\rm up}=M_A=M_B$ and $M_{\rm down}=M_C$ 
up to $n_{\rm max}=10$ for $s=1/2$ and $1$ and  up to $n_{\rm max}=8$ for
$s=3/2,2,\dots,4$.
Again we perform an extrapolation of the CCM-SUB$n$-$n$ data applying
$M_{X}=b_0+b_1(1/n)+b_2(1/n)^{2}$, $X=A,B,C$.
We use only even SUB$n$-$n$
approximations  for the  extrapolation and the corresponding scheme is
therefore 'extra1'
(see Sec.~\ref{ccm}). 
The resulting  data for $M_A=M_B$ and $M_C$ are given in Table~\ref{table4}.
We find that $M_{\rm up}$ is always larger than $|M_{\rm down}|$. In particular,  
the difference in the magnitude of both quantities is remarkably
large for $s=1/2$.
As for the zero-field case, the sublattice
magnetizations within the  plateau state are reduced  by quantum
fluctuations.
This reduction is,
however, much smaller than that for the
canted zero-field state.

\begin{figure}[ht]%[ht]  h!
\begin{center}
\epsfig{file=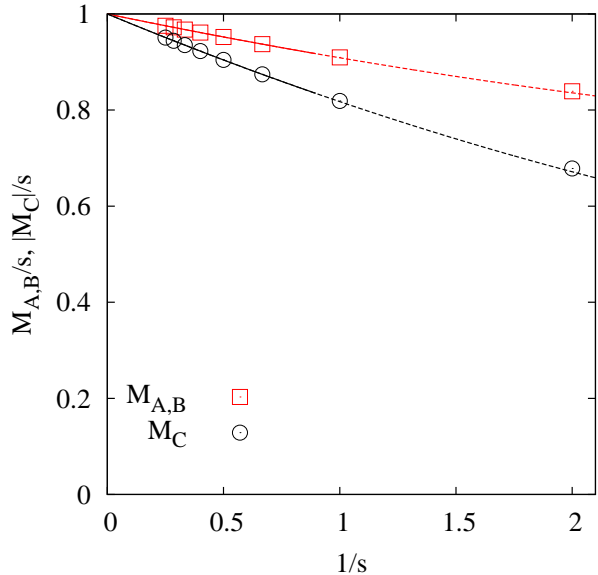,scale=0.9,angle=0.0}
\end{center}
\caption{
Extrapolated CCM data (scheme 'extra1',
see main text) for the sublattice magnetizations in the plateau state,
$M_{A}= M_{B}$ and $M_C$, as a function of  $1/s$.
The symbols represent the CCM data points, the
corresponding red
and black lines show the fit function
$M_X(s)=M_X|_{s\to\infty}\left(1-x_1/s-x_2/s^2\right)$.
 }
\label{M_ABC_vs_s}
\end{figure}

We obtain the $1/s$ dependence for $M_A=M_B$ and $M_C$   
by fitting  our extrapolated CCM data  for $s \ge 3/2$
   using (as previously)  the ansatz
$X(s)=X|_{s\to\infty}\left(1-x_1/s-x_2/s^2\right)$, $X=M_{A,B,C}$.
The classical values are
$M_{A,B}|_{s\to\infty}=s$ and $M_{C}|_{s\to\infty}=-s$.
The values for $1/s$ expansion parameters $x_1$ and $x_2$ are 
$x_1=-0.1003$ and $x_2=0.0091$ for $M_{A,B}$ and $x_1=+0.2006$ and
$x_2=-0.0182$ for $M_{C}$ and  the corresponding $1/s$ behavior is shown in Fig.~\ref{M_ABC_vs_s}.

\begin{table}[htb]
    \centering
\caption{Extrapolated CCM results (extrapolation scheme 'extra1') for 
the  sublattice
magnetizations in the plateau state, $M_{X}\vert_{n \to
\infty}$, $X=A,B,C$.
}
$ $\\
    \begin{tabular}{lrrrrrrrr}\hline\hline
      &\; $s=1/2$ &\; $s=1$&\; $s=3/2$ &\; $s=2$ &\; $s=5/2$ &\; $s=3$ &\;  $s=7/2$ &\;  $s=4$   \\ \hline
$M_{A,B}$  &\;   0.8392   &\;  0.9095   &\;   0.9372  &\;    0.9521  &\;  0.9614    &\;   0.9676  &\;  0.9721 &\; 0.9755 \\   
$M_C$     &\;  -0.6783     &\;   -0.8190    &\;   -0.8743   &\;   -0.9043   &\;   -0.9227 &\;   -0.9352 &\; -0.9441 &\; -0.9509 
    \\\hline \hline
 \end{tabular}
\label{table4}
\end{table}

\begin{figure}[ht]%[ht]  h!
\begin{center}
\epsfig{file=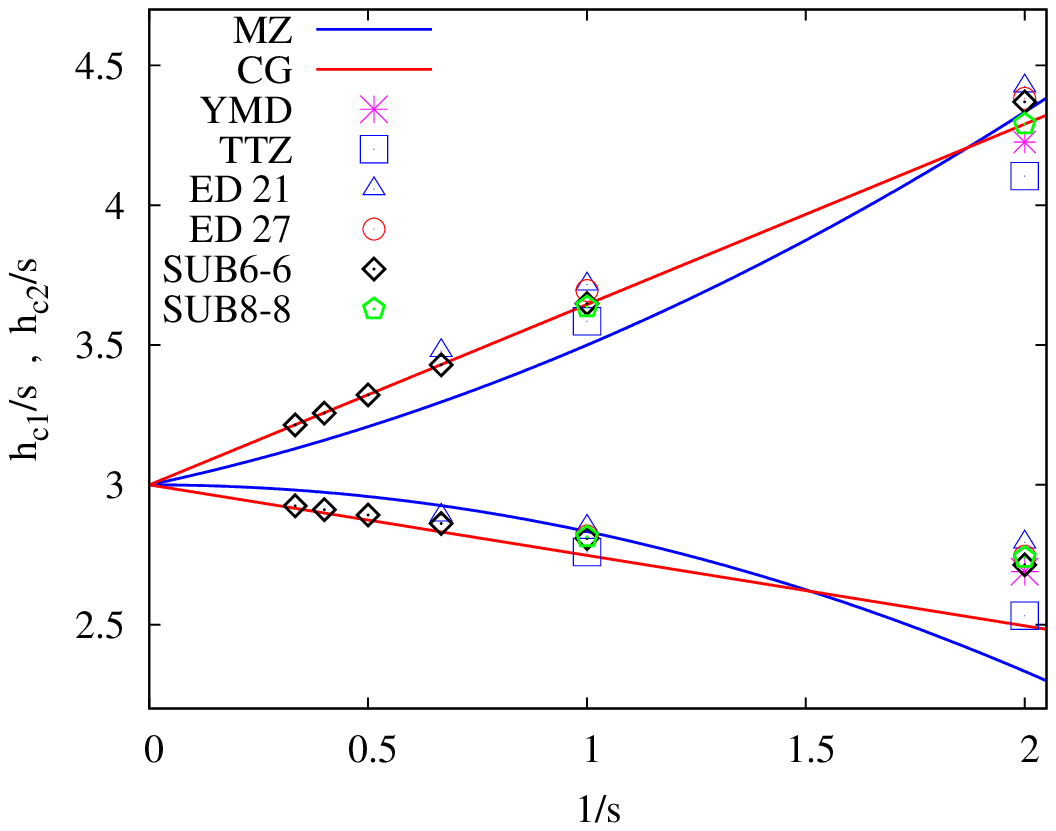,scale=0.9,angle=0.0}
\end{center}
\caption{Critical fields, $h_{c1}$ and $h_{c2}$, that bound  the  $1/3$
plateau versus the inverse of the spin quantum number, $1/s$, calculated  by exact diagonalization for
$N=21$ and $27$ (labels ED 21 and ED 27) 
and CCM (labels SUB4-4 and SUB6-6) compared to large-$s$ approaches \cite{chub_gol,zhito2011,zhito2014} and cluster mean-field
data \cite{danshita2014}. The  blue lines (labeled by MZ) correspond to 
$h_{c1}/s= 3 -1/6s^2$ and $h_{c2}/s= 3+1/3s+1/6s^2$ 
(Ref.~\cite{zhito2014}) and the red lines (labeled by CG) correspond to
$h_{c1}/s= 3-0.028/s$ and $h_{c2}/s= 3+0.0717/s$ 
(Ref.~\cite{chub_gol}).
The abbreviations YMD and TTZ represent  data from Ref.~\cite{danshita2014} and
Ref.~\cite{zhito2011}, respectively.
 }
\label{plateau_ends}
\end{figure}

We shall now discuss  in some detail the  critical fields, $h_{c1}$ and $h_{c2}$,
that define the position and the width of the plateau in the $M(h)$ curve. 
These results are also relevant for the experimental searches of the
magetization plateau in magnetic compounds with triangular-lattice
structure, cf. 
Refs.~\cite{shirata2011,shirata2012,tanaka2013a,wir_und_tanaka2013}.  We
present our data in Table~\ref{table_hc}
and show the $1/s$ dependences of
$h_{c1}$ and $h_{c2}$ in Fig.~\ref{plateau_ends}.  
In addition to our CCM and
ED data, we show also relevant data for $h_{c1}$ and $h_{c2}$
from Refs.~\cite{chub_gol,zhito2011,zhito2014,danshita2014} for comparison. 
The monotonic shrinking of the plateau widths known for the large-$s$
approaches is also clearly seen in our CCM and ED data.
    
We notice that the plateau ends at $h_{c1}$ and $h_{c2}$ behave
differently with increasing $s$.  Although the lower plateau end is only
slightly shifted, the shift of the upper one at $h_{c2}$ is more pronounced,
see also Figs.~\ref{m_h}-\ref{angles}. 
We also see that the data for  $h_{c1}$ and $h_{c2}$ provided in the literature
for the extreme quantum case $s=1/2$ exhibit a rather large amount of scattering. 
As already found for the zero-field sublattice magnetization and
susceptibility,
cf.  Sec.~\ref{H_0}, the quasi-classical large-$s$
approaches~\cite{chub_gol,zhito2011,zhito2014} for $s=1/2$ noticeably
deviate from our data directly calculated for $s=1/2$.
Moreover, the recent real-space perturbation theory\cite{zhito2014} yields
an $1/s$-dependence of $h_{c1}$ and $h_{c2}$ that  
significantly deviates from the spin-wave, ED  as well as the CCM behavior.
On the other hand,  
the recent large-scale cluster mean-field
approach of Ref. \cite{danshita2014} yields $h_{c1}/s=2.690$ and 
$h_{c2}/s= 4.226$ for the $s=1/2$ case, and these values are close
to the CCM-SUB8-8 results.

%\begin{widetext}
\begin{table}[htb]
    \centering
\caption{Critical  fields
 $h_{c1}$ and  
$h_{c2}$, where the one-third plateau begins and ends. Note that for $s=1/2$
 ED data also for $N=36$ and $39$ are available \cite{Hon1999,HSR04,Sakai}.
}
$ $\\ 
    \begin{tabular}{ccccccccc}\hline\hline
    \parbox[0pt][1.5em][c]{0cm}{}        & \multicolumn{8}{c} {CCM} \\ \hline
    \parbox[0pt][1.5em][c]{0cm}{}        & \multicolumn{2}{c} { SUB2-2}& \multicolumn{2}{c} { SUB4-4} &  \multicolumn{2}{c} { SUB6-6} &  \multicolumn{2}{c} {  SUB8-8} \\ \hline
   $s$ &   \; $h_{c1}/s$ &\; $h_{c2}/s$ &\; $h_{c1}/s$ &\; $h_{c2}/s$&\; $h_{c1}/s$ &\; $h_{c2}/s$&\; $h_{c1}/s$ &\; $h_{c2}/s$  \\ \hline
    $1/2$      &\;  2.382  &\; 4.344     &\;  2.624  &\; 4.482        &\; 2.714  &\; 4.370  &\; 2.740  &\; 4.290          \\
    $1$        &\;  2.727  &\; 3.617     &\;  2.788  &\; 3.674        &\; 2.809  &\; 3.648  &\; 2.814  &\; 3.637          \\
    $3/2$      &\;  2.826  &\; 3.405     &\;  2.854  &\; 3.440        &\; 2.862  &\; 3.429  &\;    --  &\; --             \\
    $2$        &\;  2.873  &\; 3.303     &\;  2.888  &\; 3.327        &\; 2.892  &\; 3.322  &\;    --  &\; --             \\
    $5/2$      &\;  2.900  &\; 3.242     &\;  2.909  &\; 3.261        &\; 2.911  &\; 3.257  &\;    --  &\; --             \\
    $3$        &\;  2.917  &\; 3.202     &\;  2.923  &\; 3.217        &\; 2.925  &\; 3.214  &\;    --  &\;    --            \\\hline \hline
    \parbox[0pt][1.5em][c]{0cm}{}       &  \multicolumn{8}{c} {ED} \\\hline
    \parbox[0pt][1.5em][c]{0cm}{}   & \multicolumn{2}{c} { $N=12$}& \multicolumn{2}{c}    { $N=18$} &  \multicolumn{2}{c}     {$N=21$} &  \multicolumn{2}{c}   {$N=27$} \\\hline
   $s$ &   \; $h_{c1}/s$ &\; $h_{c2}/s$ &\; $h_{c1}/s$ &\; $h_{c2}/s$ &\; $h_{c1}/s$ &\; $h_{c2}/s$  &\; $h_{c1}/s$ &\; $h_{c2}/s$ \\ \hline
    $1/2$              &\;  2.615 &\; 4.934     &\;  2.805 &\; 4.578         &\; 2.794   &\; 4.425    &\;  2.745  &\; 4.382      \\
    $1$                &\;  2.816 &\; 3.916     &\;  2.879 &\; 3.786         &\; 2.851   &\; 3.717    &\;  2.817  &\; 3.695      \\
    $3/2$              &\;  2.864 &\; 3.613     &\;  2.913 &\; 3.523         &\; 2.890   &\; 3.479    &\;   --    &\;   --       \\
    $2$                &\;  2.893 &\; 3.461     &\;  2.932 &\; 3.391         &\; --      &\;   --     &\;   --    &\;   --       \\
    $5/2$              &\;  2.911 &\; 3.369     &\;    --  &\;  --           &\; --      &\;   --     &\;   --    &\;   --       \\
    $3$                &\;  2.924 &\; 3.308     &\;    --  &\;  --	          &\; --      &\;   --     &\;   --    &\;   --       \\\hline \hline
 \end{tabular}
\label{table_hc}
\end{table}
%\end{widetext}

\vspace{1cm}

\section{Summary}

The HAFM
on the triangular
lattice is a basic model of quantum magnetism. 
The theoretical treatment of this frustrated spin model is challenging and the
precision of the existing data 
for the basic parameters determining  the low-energy physics 
of the model 
is less than that of corresponding unfrustrated
models such as the square-lattice HAFM.
Furthermore several magnetic compounds exist, see e.g.
Refs.~\cite{shirata2011,shirata2012,tanaka2013a,Nakatsuji2007,Reotier2007,Bao2009,Reotier2015}, that are described by this
model. Hence, there is a need to improve the accuracy of the data available
from theoretical investigations  also from the experimental
point of view.

In the present paper we present        
large-scale CCM calculations for the basic GS parameters,
energy $e_0$, sublattice magnetization  $M_{\rm sub}$, in-plane spin
stiffness $\rho_s$ and 
in-plane magnetic susceptibility $\chi$, for arbitrary spin quantum number
$s$.
In addition to these zero-field quantities, we also consider the magnetization
process.
 It is known from many previous studies for other frustrated quantum spin models, such
as the kagome HAFM and the $J_1$-$J_2$ square-lattice HAFM, that the 
CCM provides accurate GS results. Hence, the results presented  here will
contribute to 
improve available theoretical data, especially for the extreme quantum cases  
$s=1/2$ and $s=1$ where 
large-$s$ (spin-wave) theories do not to provide sufficently precise data.

%\textcolor{red}{
Although the present study is purely theoretical, the data presented here
might be used to interpret experimental results for  magnetic compounds 
that are described by the triangular-lattice HAFM, see e.g.
Refs.~\cite{shirata2011,shirata2012,tanaka2013a,Nakatsuji2007,Reotier2007,Bao2009,Reotier2015}. 
As already mentioned above, the compounds Ba$_3$CoSb$_2$O$_9$ and
Ba$_3$NiSb$_2$O$_9$ are described well by the Heisenberg model
considered here with spin quantum number $s=1/2$
(Ba$_3$CoSb$_2$O$_9$)\cite{shirata2012} and $s=1$
(Ba$_3$NiSb$_2$O$_9$)\cite{shirata2011,wir_und_tanaka2013}. 
We remark that very good agreement between the theoretical CCM data 
and experimental data has been reported for these cases.
The CCM data presented here for $s=3/2$ and $s=2$ might be useful 
for further studies of the magnetic compounds La$_2$Ca$_2$MnO$_7$ 
(with $s=3/2$) \cite{Bao2009,Reotier2015} and FeGa$_2$S$_4$  
(with $s=2$) \cite{Nakatsuji2007,Reotier2007}.


\begin{thebibliography}{200}
\bibitem{defin_chi} $\chi$ is sometimes defined per volume, see e.g.
Ref.~\cite{chub94}, which yields a factor of $2/\sqrt{3}$ for the triangular lattice.
\bibitem{SUBn-n} Note that  the notation LSUB$n$ is used  for $s=1/2$  typically,
see, e.g., Ref.~\cite{farnell2009}. Thus we
mention here that for $s=1/2$ the LSUB$n$ scheme is
identical to the SUB$n$-$n$ scheme.
\bibitem{comp_archi} We notice that in Ref.~\cite{archimedean2014}, where the
triangular lattice for $s=1/2$ was compared with other  Archimedean
lattices,  
we used  SUB$n$-$n$ data for $n=4,5,\ldots,10$ for the extrapolation.
This was carried out in order  to have consistency within the class of Archimedean-lattice
antiferromagnets with non-collinear ground states.
\end{thebibliography}
\end{document}